\documentclass[aps,twocolumn,superscriptaddress,notitlepage,nofootinbib,showkeys,showpacs]{revtex4-1}

\usepackage{epsfig}
\usepackage{multirow}
\usepackage{slashed}
\usepackage{amsmath}
\usepackage{physics}
\usepackage{qcircuit}
\usepackage{braket}
\usepackage{graphicx}
\usepackage{subfigure}
\usepackage{ulem}
\usepackage{color}
\usepackage{epstopdf}
\usepackage[inline]{enumitem}
\usepackage{amssymb}

\usepackage{ulem}

\def\bea#1\eea{\begin{align}#1\end{align}}
\newcommand{\nn}{\nonumber\\}
\newcommand{\bef}{\begin{figure}[!htp]}
\newcommand{\eef}{\end{figure}}

\begin{document}

\title{Exploring Light-Cone Distribution Amplitudes from Quantum Computing}

\date{\today}

\author{Tianyin Li}
\affiliation{Guangdong Provincial Key Laboratory of Nuclear Science, Institute of Quantum Matter, South China Normal University, Guangzhou 510006, China}
\affiliation{Guangdong-Hong Kong Joint Laboratory of Quantum Matter,
Southern Nuclear Science Computing Center, South China Normal University, Guangzhou 510006, China}

\author{Xingyu Guo}
\affiliation{Guangdong Provincial Key Laboratory of Nuclear Science, Institute of Quantum Matter, South China Normal University, Guangzhou 510006, China}
\affiliation{Guangdong-Hong Kong Joint Laboratory of Quantum Matter, Southern Nuclear Science Computing Center, South China Normal University, Guangzhou 510006, China}

\author{Wai Kin Lai}
\email{wklai@m.scnu.edu.cn}
\affiliation{Guangdong Provincial Key Laboratory of Nuclear Science, Institute of Quantum Matter, South China Normal University, Guangzhou 510006, China}
\affiliation{Guangdong-Hong Kong Joint Laboratory of Quantum Matter,
Southern Nuclear Science Computing Center, South China Normal University, Guangzhou 510006, China}
\affiliation{Department of Physics and Astronomy, University of California, Los Angeles, California 90095, USA}

\author{Xiaohui Liu}
\affiliation{Center of Advanced Quantum Studies, Department of Physics,
Beijing Normal University, Beijing 100875, China}
\affiliation{Center for High Energy Physics, Peking University, Beijing 100871, China}

\author{Enke Wang}
\affiliation{Guangdong Provincial Key Laboratory of Nuclear Science, Institute of Quantum Matter, South China Normal University, Guangzhou 510006, China}
\affiliation{Guangdong-Hong Kong Joint Laboratory of Quantum Matter,
Southern Nuclear Science Computing Center, South China Normal University, Guangzhou 510006, China}

\author{Hongxi Xing}
\email{hxing@m.scnu.edu.cn}
\affiliation{Guangdong Provincial Key Laboratory of Nuclear Science, Institute of Quantum Matter, South China Normal University, Guangzhou 510006, China}
\affiliation{Guangdong-Hong Kong Joint Laboratory of Quantum Matter,
Southern Nuclear Science Computing Center, South China Normal University, Guangzhou 510006, China}

\author{Dan-Bo Zhang} 
\affiliation{Guangdong-Hong Kong Joint Laboratory of Quantum Matter, Frontier Research Institute for Physics,
South China Normal University, Guangzhou 510006, China}  
\affiliation{Guangdong Provincial Key Laboratory of Quantum Engineering and Quantum Materials, School of Physics and Telecommunication Engineering,
South China Normal University, Guangzhou 510006, China}

\author{Shi-Liang Zhu} 
\affiliation{Guangdong-Hong Kong Joint Laboratory of Quantum Matter, Frontier Research Institute for Physics,
South China Normal University, Guangzhou 510006, China}  
\affiliation{Guangdong Provincial Key Laboratory of Quantum Engineering and Quantum Materials, School of Physics and Telecommunication Engineering,
South China Normal University, Guangzhou 510006, China}

\collaboration{QuNu Collaboration}

\date{\today}         

\begin{abstract}
Light-cone distribution amplitudes (LCDAs) are essential nonperturbative quantities for theoretical predictions of exclusive high-energy processes in quantum chromodynamics (QCD). We demonstrate the prospect of calculating LCDAs on a quantum computer by applying a recently proposed quantum algorithm,
with staggered fermions, to the simulation of the LCDA in the (1+1)-dimensional Nambu-Jona-Lasinio (NJL) model on classical hardware. The agreement between the result from the classical simulation of the quantum algorithm and that from exact diagonalization justifies the proposed quantum algorithm. We find that the resulting LCDA in the NJL model exhibits features shared with the LCDAs obtained from QCD.
\end{abstract}



\maketitle

\section{Introduction}
Light-cone distribution amplitudes (LCDAs) are quantities that describe the nonperturbative aspects of bound states in high-energy exclusive processes in quantum chromodynamics (QCD). They are complementary to the parton distribution functions (PDFs), which are nonperturbative quantities for inclusive QCD processes with incoming hadrons, such as deep-inelastic scattering of the proton at the Hadron-Electron Ring Accelerator (HERA) and future Electron-Ion Colliders ~\cite{H1:2015ubc,AbdulKhalek:2021gbh,EicCWP,Anderle:2021wcy}.
In an exclusive high-energy QCD process, up to corrections suppressed by inverse powers of a large energy scale, the scattering amplitude can be factorized
into the convolution of a perturbative Wilson coefficient and a nonperturbative amplitude, the LCDA~\cite{Lepage:1980fj,Efremov:1979qk}. A seminal example is the electromagnetic form factor $F(Q^2)$ for the process $\gamma^*\gamma\to q{\bar q} \to \pi^0$ at large momentum transfer $Q$, for which the factorization reads
\begin{align}
F(Q^2)=f_\pi\int_0^1dx\, T_H(x,Q^2;\mu)\phi_\pi(x;\mu)
+ \mathcal{O}(\Lambda_{\rm QCD}^2/Q^2)  \,, 
\end{align}
where $Q^2$ is the 4-momentum squared carried by the virtual photon $\gamma^*$, $\Lambda_{\rm QCD}$ is the energy scale below which QCD becomes nonperturbative, and $\mu$ is the factorization scale, which separates the short-distance physics from the long-distance wave function $\phi_\pi$.
Here, $T_H(x,Q^2;\mu)$ is the hard kernel, which describes the production of a quark-antiquark pair by
short-distance dynamics. $T_H(x,Q^2;\mu)$ is perturbatively calculable as an expansion in the strong coupling constant $\alpha_s$.
The wave function $ \phi_\pi(x,\mu)$ and the factor $f_\pi$ are the LCDA and the decay constant of the neutral pion, respectively. They together encode the hadronization of a quark-antiquark pair into a pion,
and are nonperturbative as they are sensitive to the long-distance dynamics of QCD. The LCDA of a meson can be viewed as the probability amplitude 
to find the valence $q\bar{q}$ Fock state  
in which the quark $q$ and the antiquark ${\bar q}$ carry respectively the momentum fraction $x$ and $1-x$ of the highly boosted meson. The decay constant is defined as the overall normalization of the LCDA.
The LCDAs and decay constants for baryons can be likewise defined.
The LCDAs, being the essential ingredients for reliable
predictions for exclusive QCD processes, have been studied intensively in various directions.
Early studies include their perturbative evolution with the scale $\mu$ and their asymptotic behaviors~\cite{Lepage:1980fj,Efremov:1979qk,Chernyak:1983ej}.
Estimations of the LCDAs using 
sum rules, illustrative models,~\cite{Chernyak:1981zz,Chernyak:1984bm,King:1986wi,Chernyak:1987nu,Chernyak:1987nv,Radyushkin:1990te,RuizArriola:2006jge,
RuizArriola:2002bp,Chang:2013pq,Agaev:2012tm}, light-front quantization~\cite{Brodsky:2006uqa,Vary:2009gt,Brodsky:2014yha,Xu:2021wwj}, as well as the refactorization in the nonrelativistic expansion~\cite{Ma:2006hc,Jia:2008ep} have also been discussed under various circumstances in the literature.
Evaluations of the LCDAs in lattice QCD were initially performed by taking the 
moments~\cite{ Gockeler:2008xv,QCDSF:2008qtn,Braun:2014wpa,Bali:2015ykx,RQCD:2019osh,Zhang:2020gaj}, and later with direct calculations in the momentum fraction $x$ within the large momentum effective theory (LaMET) framework~\cite{Zhang:2017bzy,Zhang:2017zfe,Hua:2020gnw,Hua:2022kcm}.
There are two major obstacles to acquiring  knowledge about the LCDAs, the first being the elusiveness of the relevant experimental data that hinders QCD global analyses, the second being the nature of real-time evolution on the light cone involved in the definition of LCDAs, which is not amenable to direct evaluations using Euclidean lattice QCD~\cite{Alexandru:2016gsd}.
 
Stimulated by the promising prospect of quantum computing~\cite{Arute:2019zxq}, there has been a rapidly growing
wave of research
on applications of quantum computing in elementary particle physics~\cite{Cloet:2019wre,Zhang:2020uqo,Bauer:2022hpo}.
Early theoretical attempts had showed that quantum computation costs polynomial time in simulations of 
real time dynamics in quantum field theory~\cite{Jordan:2012xnu,Jordan:2011ci,Jordan:2014tma,Klco:2018zqz}.
Validity of quantum computing in various problems in
particle physics have been studied by proposals of quantum algorithms as well as simulations 
with real quantum computers or classical hardware.
These studies include 
evaluations of nonperturbative quantities 
~\cite{Dumitrescu:2018njn,Lu:2018pjk,Lamm:2019uyc,Mueller:2019qqj,Roggero:2019myu,Echevarria:2020wct,Kreshchuk:2020kcz,Bauer:2021gup,Atas:2021ext,Li:2021kcs,Gallimore:2022hai},
simulations of real-time processes
~\cite{Martinez:2016yna,Hu:2019hrf,Bauer:2019qxa,DeJong:2020riy,Zhou:2021kdl,deJong:2021wsd,Williams:2021lvr,Atas:2022dqm,Yao:2022eqm}, as well as evaluations of thermodynamical quantities at finite chemical potential~\cite{Czajka:2021yll,Xie:2022jgj}.

Recently, a quantum algorithm was proposed for both the preparation of a hadron state and the evaluation of real-time light-like correlators in Ref.~\cite{Li:2021kcs}. The algorithm was demonstrated feasible by evaluating directly the parton distribution function in the (1+1)-dimensional Nambu-Jona-Lasinio (NJL) model with staggered fermions on classical hardware. The results obtained from the quantum algorithm were checked against exact diagonalization to find full agreements, which justifies the validity of the algorithm and suggests the possibility of evaluating the hadron parton distributions in QCD by quantum computation.

In this study, we extend the previous studies to apply the quantum algorithm to study the LCDA in the $(1+1)$D $1$-flavor NJL model, using staggered fermions. In Sec.~\ref{def_lcda}, we provide the operator definition of the LCDA in the (1+1)D NJL model. Then we present the quantum algorithm for both the hadronic state preparation and the direct computation of the quark-antiquark correlator in Sec.~\ref{algrm}. By implementing the proposed algorithm on classical hardware, we found good consistency between the results obtained from the quantum algorithm and exact diagonalization. The final results for the LCDA are presented in Sec.~\ref{res}. We give a summary in Sec.~\ref{sum}.

\section{Light-cone distribution amplitude and the NJL model}\label{def_lcda}

The LCDA of a meson $h$ is defined as
\bea\label{eq:Gdef}
	\phi_h(x) =\frac{1}{f_h}&\int {dz}\,e^{-i(x-1)n\cdot P  z}
	\nn 
&\times \bra{\Omega}\bar{\psi}(z n)\gamma^+ W(zn,0)\psi(0)\ket{h(P)} \,, 
\eea
where $P$ is the momentum of the meson and $n$ is a lightlike vector defined by $n = (1,-\hat{\mathbf{n}})$, with $\hat{\mathbf{n}}$ being a spatial unit vector along the direction of motion of the meson. The plus-component of the gamma matrix $\gamma^\mu$ in light-cone coordinates is denoted by $\gamma^+$,
i.e. $\gamma^+=n\cdot \gamma$.
The prefactor $f_h$ is the decay constant, defined such that $\int_0^1dx\,\phi_h(x)=1$. The matrix element in the second line in eq.~(\ref{eq:Gdef}) describes the transition amplitude from the vacuum $|\Omega\rangle$ to the hadron state $|h(P)\rangle$ via insertion of a quark-antiquark-pair operator. The $W(zn,0)$ is the Wilson line (gauge link) on the light cone,
\bea
	 W(zn,0)=\mathcal{P}\exp(ig\int_0^{z}dz' A^+_a(z'n)t_a)\,, \label{eq:Wilson_line}
\eea
where $t_a$ are the $\rm{SU}(3)$ fundamental generators, $\mathcal{P}$ denotes path-ordering, and $A^+_a=n\cdot A_a$ is the 
plus-component of the $\rm{SU}(3)$ gauge potential. Generally, to simulate the LCDA on a quantum computer, one has to: 
\begin{enumerate*} 
\item prepare the vacuum state $\ket{\Omega}$ and the hadronic state $\ket{h(P)}$ on the quantum computer; \item simulate the Wilson line on the quantum computer; 
\item evaluate the matrix element $\bra{\Omega}\mathcal{O}\ket{h}$.
\end{enumerate*} 
We will elaborate on the first and third steps later in Section~\ref{algrm}. 
We note that the presence of the gauge field in a gauge theory will dramatically increase the demand for quantum resources. However, as noted in ref.~\cite{Li:2021kcs}, simulating the Wilson line in the second step does not introduce substantial complexity even though the Wilson line is non-local, which is contrary to the claim in ref.~\cite{Lamm:2019uyc}.

Since simulating QCD on a quantum computer remains a formidable task to date~\cite{Banuls:2019bmf}, we resort to a simple model,
the Nambu-Jona-Lasinio (NJL) model~\cite{Nambu:1961tp,Nambu:1961fr} in (1+1) dimensions, 
also known as the Gross-Neveu model~\cite{Gross:1974jv}, in order to demonstrate simulations of the LCDA on a quantum computer. The Lagrangian of the $(1+1)$D 
NJL model is given by
\begin{equation}\label{eq:LCDA}
	\mathcal{L}=\bar{\psi} (i\gamma^\mu \partial_\mu-m_q)\psi +g(\bar{\psi} \psi)^2\,,
\end{equation}
where $g$ is the strong coupling constant and $m_q$ is the quark mass. 
The LCDA $\phi_h(x)$ of a meson $h$ in the NJL model is defined by
\bea\label{eq:def}
	\phi_h(x) =\frac{1}{f_h}&\int {dz}\,e^{-i(x-1)n\cdot P  z}
	\nn 
&\times \bra{\Omega}\bar{\psi}(z n)\gamma^+ \psi(0)\ket{h(P)} \,.
\eea 
We note that the LCDA defined as in eq.~(\ref{eq:def}) is independent of the frame of reference.
For the sake of practical computation, we will evaluate the LCDA in the rest frame of the meson, in which case we have
\bea\label{eq:LCDA}
	\phi_h(x) =\frac{1}{f_h}&
	\int {dz}\,e^{-i(x-1)m_h  z}
	\nn 
&\times 
\bra{\Omega} e^{iHt}\bar{\psi}(0,-z)e^{-iHt}\gamma^+ \psi(0,0)\ket{h} \,,  
\eea 
where we have written the quark field ${\bar \psi}(zn)$ as ${\bar \psi}(zn) = e^{iHz}\bar{\psi}(0,-z) e^{-iHz}$ with $H$ the Hamiltonian of the NJL model, and we will set $t = z$ in eq.~(\ref{eq:LCDA}) to put the correlator on the light cone.
Here $m_h$ is the mass of the meson $h$. 

To facilitate quantum simulations, we discretize the space into $N/2$ lattice sites and place the fermion field on the lattice following:
\bea
\psi(0,\mathbf{z}) =  
\begin{pmatrix}
\psi_{1}(0,\mathbf{z}) \\
\psi_{2}(0,\mathbf{z}) 
\end{pmatrix}	\equiv
\begin{pmatrix}
\varphi_{2n} \\
\varphi_{2n+1} 
\end{pmatrix}
\,, 
\eea
where $0 \le n 
\le \frac{N}{2}-1$. Notice that, throughout this paper, the subscript $n$ denotes the qubit index. Note that we have distributed the upper and lower component of the Dirac spinor to the even and odd lattice sites respectively. 
After performing the Jordan-Wigner transformation \cite{backens_shnirman_makhlin_2019}, 
\bea
\varphi_{n}  
\equiv 
\Xi_{n}^3
\sigma_{n}^+
\,, 
\eea
the fields operator $\varphi_n$ can be represented by quantum gates on a quantum circuit. 
Here we have introduced
the raising and lowering operators $\sigma_{n}^{\pm} = \frac{1}{2}( \sigma_{n}^1 \pm i \sigma_{n}^2)$, and the string operator $\Xi_{n}^3\equiv \prod_{n'<n}\sigma^3_{n'}$, with $\sigma^{j}_{n}$ denoting the $j$-th component of the Pauli matrix on the qubit $n$. Throughout this study, we impose the periodic boundary condition.
The LCDA then reads
\bea
		\phi_h(x)&=\frac{1}{f_h}\sum_z \frac{1}{4\pi}e^{-i(x-1)m_h z}\tilde\phi_h(z)\,,  \label{eq:Fourier_decomp}
\eea
where
\bea\label{eq:def_D}
		\tilde\phi_h(z)&=\sum_{i,j=0}^{1}
		\bra{\Omega} e^{iHz}
        \varphi^\dagger_{-2z+i}
		e^{-iHz}
        \varphi_{j} 
\ket{h}\,.
\eea
\section{Quantum algorithm for LCDAs}
\label{algrm}
\begin{figure}[htbp]
	\centering
	\subfigure[Quantum circuit for preparation of the hadronic state]{
	\label{QAOA-circuit}
	\includegraphics[width=0.35 \textwidth]{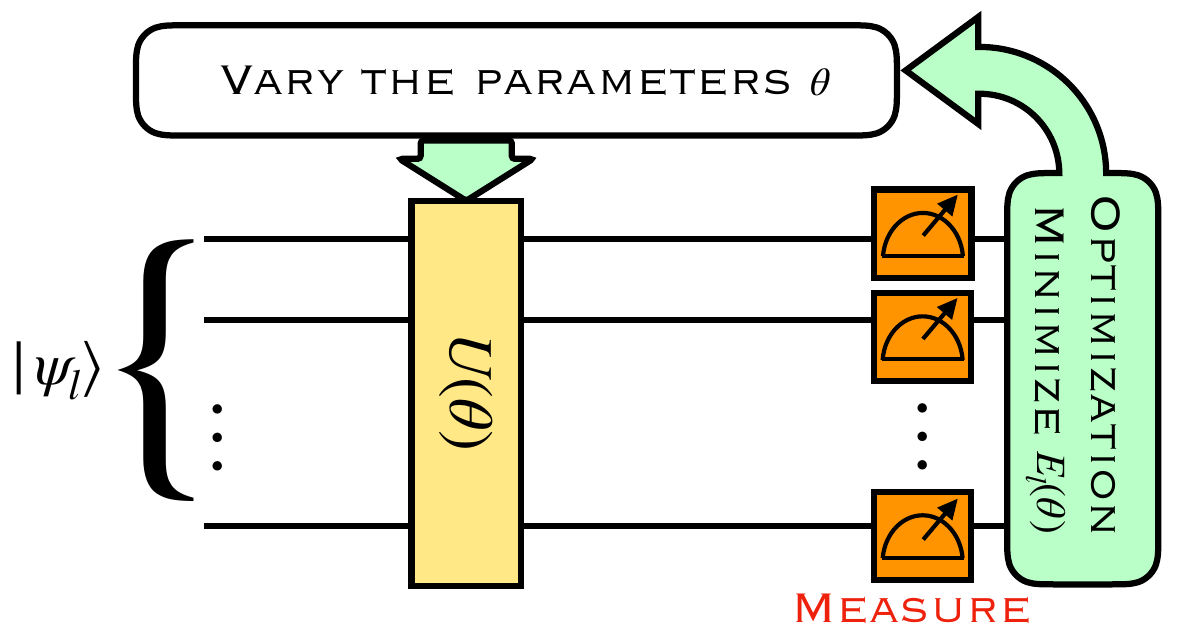}}
	\subfigure[Quantum circuit for the correlator $\bra{\Omega}\mathcal{O}\ket{h}$]{
	\label{corr-circuit}
	\includegraphics[width=0.48 \textwidth]{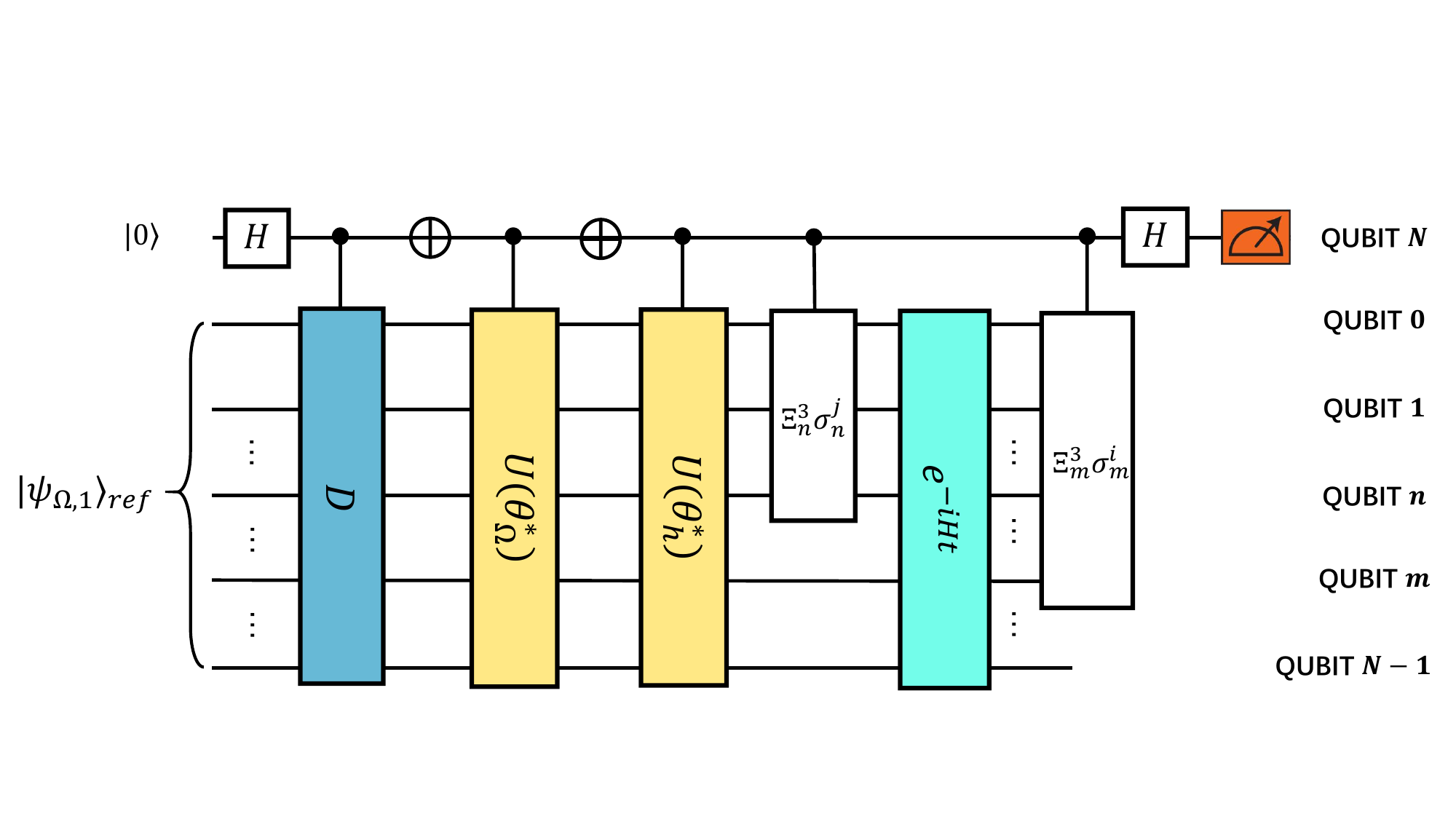}}
	\caption{Quantum circuit for calculation of the LCDA. Diagram (a) gives the circuit for hadronic state preparation, while (b) is for evaluation of the correlator. }
	\label{fig-circuit}
\end{figure}

We implement the quantum algorithm proposed in Ref.~\cite{Li:2021kcs} to simulate the real-time light-cone correlator.
The quantum algorithm is described by the quantum circuit shown in Fig.~\ref{fig-circuit}, which consists of two parts:
preparation of the hadronic state and evaluation of the correlator. Note that the vacuum state can be prepared as
a particular case of a hadronic state.

\subsection{Preparation of the hadronic state}
The preparation of the hadronic state is achieved by the quantum-number-resolving variational quantum eigensolver (VQE), summarized as follows.
To find the first $k$ excited states $\ket{h}$ with quantum numbers $l$, we contruct the trial states $\ket{\psi_{li}(\theta)}$, $i=1,2,\dots,k$, ($i=1$ corresponds to the lowest-lying state) by 
\begin{equation}
\ket{\psi_{li}(\theta)}
= U(\theta) \ket{\psi_{li}}_{\text{ref}}\,,
\end{equation}
where $\ket{\psi_{li}}_{\text{ref}}$ are some input reference states with the same quantum numbers $l$ as the hadron states $\ket{h}$, and $U(\theta)$ a symmetry-preserving unitary operator with parameters $\theta$. Since $U(\theta)$ preserves the quantum numbers, all of the trial states will have the same quantum numbers as $\ket{h}$. Then we can find out the hadron states among the trial states by minimizing the cost function 
\begin{equation}
	E_l(\theta)=\sum_{i=1}^k w_{li} \bra{\psi_{li} (\theta)}H\ket{\psi_{li} (\theta)} \,.
\end{equation}
Here we require $w_{l1}>w_{l2}>\cdots>w_{lk}$. 
The $i$-th excited state $\ket{h}$ is then prepared as
$\ket{h} =
U(\theta^\ast)
\ket{ \psi_{li} }_{\text{ref}}$\, with $\theta^\ast$ the optimized values of $\theta$. In the following, we describe in detail
how the input reference states are prepared, and how the operator $U(\theta)$ is constructed with the quantum alternating operator ansatz (QAOA).

\subsubsection{Preparation of input reference states}
We complete the quantum algorithm framework for preparing hadronic states with detailed constructions of the input states. The input states have the same quantum numbers as the hadron and in general, they should be superposed states of the computational basis. We find that the input states are closely related to the so-called Dicke states, whose efficient preparation with explicit quantum circuits is available in Ref.~\cite{10.1007/978-3-030-25027-0_9}. To study generic hadronic states on a quantum computer, we outline the basic construction of the quantum circuit for the Dicke states. 

\begin{figure}[htbp]
	\centering
	\includegraphics[width=0.5 \textwidth]{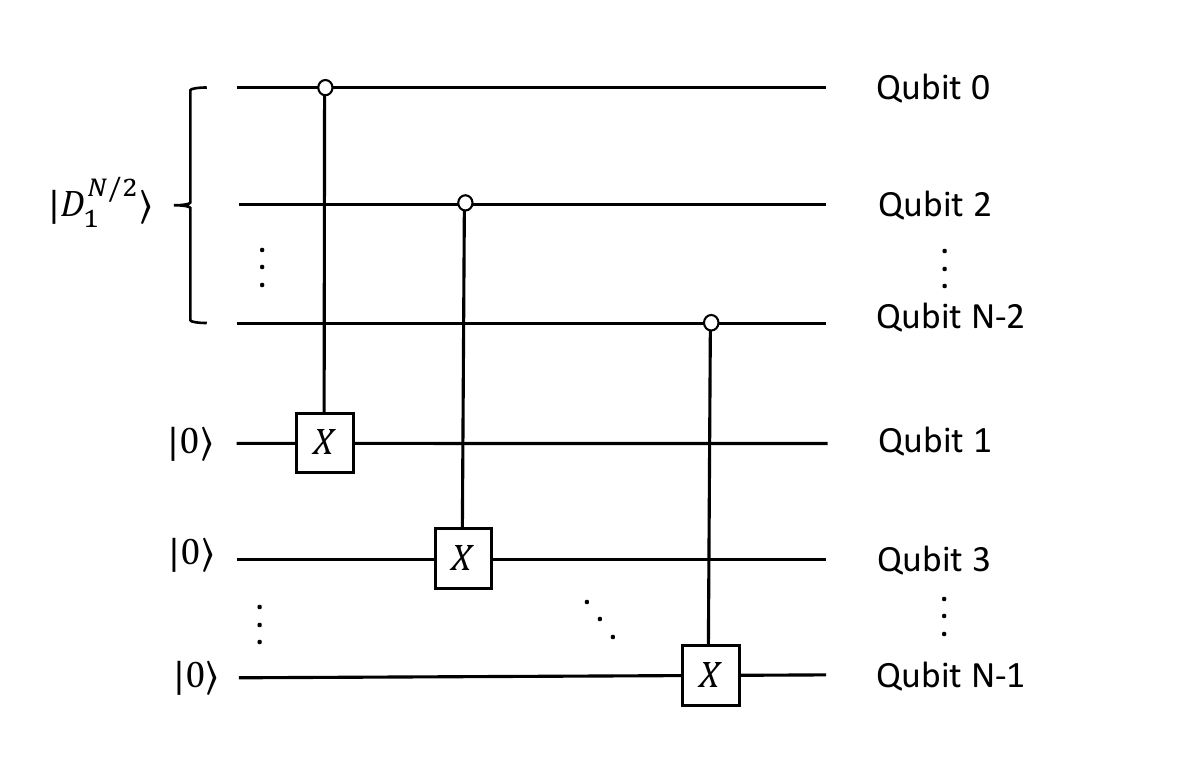}
	\caption{Quantum circuit for preparing $\ket{\psi_{\Omega,2}}_{\rm ref}$ from the Dicke state $\ket{D^{N/2}_1}$. The input state for the even qubits is the Dicke state and the input state for the odd qubits is the state $\ket{0}^{\otimes N/2}$.}	\label{fig:D-to-ini}
\end{figure}

In the NJL model, to prepare the lightest $\ket{q\bar{q}}$ state that has the same quantum numbers as the vacuum, the $N$-qubit zero-momentum input state of the QAOA can be chosen as:
    \bea\label{eq:NJLref}
        \ket{\psi_{\Omega,1}}_{\text{ref}}
        =& \ket{ 010101\dots 01} \,, \nn 
        \ket{\psi_{\Omega,2}}_{\text{ref}}
        =& \frac{1}{\sqrt{N/2}}
        \left(
        \ket{1001\dots 01} + \ket{0110\dots 01} \right.\nn
        & \left. 
        + \dots + \ket{0101\dots 10}
        \right)\,,
    \eea
where $\ket{x_1x_2...x_n}$ with $x_1,x_2,...,x_n=0,1$ are the computational basis states  for an $n$-qubit system~\cite{nielsen_chuang_2010}, both $\ket{\psi_{\Omega,1}}_{\text{ref}}$ and $\ket{\psi_{\Omega,2}}_{\text{ref}}$ share the same quantum numbers with the $\ket{q{\bar q}}$ state. 
The state $\ket{\psi_{\Omega,1}}_{\text{ref}}$ is a product state that can be easily prepared from the state $\ket{0000\dots 00}$. The preparation of the superposed state $\ket{\psi_{\Omega,2}}_{\text{ref}}$ is more involved, which we will focus on. We first denote $\ket{\bar{0}}\equiv \ket{01}$ and $\ket{\bar{1}}\equiv \ket{10}$ to write $\ket{\psi_{\Omega,2}}$ as:
\bea
    \ket{\psi_{\Omega,2}}_{\text{ref}}=\sqrt{\frac{1}{C^1_{N/2}}}(\ket{\bar{1}\bar{0}\dots \bar{0}}+\ket{\bar{0}\bar{1}\dots\bar{0}}+\ket{\bar{0}\bar{0}\dots\bar{1}})\,.
\eea
It can be seen that $\ket{\psi_{\Omega,2}}_{\text{ref}}$ is closely related to the Dicke state $\ket{D^{N/2}_1}$~\cite{10.1007/978-3-030-25027-0_9}, which can be prepared by a series of
split-and-cylic-shift ($\rm{SCS}_{n,1}$) gates as:
\bea
    \ket{D^{N/2}_1}=\prod_{n=2}^{N/2} {\rm SCS}_{n,1}\ket{0}^{\otimes \frac{N}{2}-1}\ket{1}\,.
\eea
The $\rm{SCS}_{n,1}$ gates can be written as elementary controlled NOT (CNOT) and controlled $\rm{R}_y$ gates,
\begin{align}
    &{\rm SCS}_{n,1}=\Bigg[{\rm CNOT}(n-1,n)\nonumber\\
    &\quad\times {\rm CR}_y\left(n,n-1,2\cos^{-1}\sqrt{\frac{1}{n}}\right)\nonumber\\
    &\quad \times {\rm CNOT}(n-1,n)\Bigg]\,,
\end{align}
where ${\rm CNOT}(i,j)$ is the controlled NOT gate, with the qubit $i$ being the control qubit and the NOT gate acting on the qubit $j$. ${\rm CR}_y(i,j,\theta)$ is the controlled ${\rm R}_y$ gate, where $i$ is the control qubit, and ${R}_y(\theta)$ acts on the qubit $j$ as a rotation about the $y$-axis by an angle $\theta$. 
To prepare $\ket{\psi_{\Omega,2}}_{\text{ref}}$, we first prepare the Dicke state $\ket{D^{N/2}_1}$. Then to the right of each qubit we attach a new qubit initialized as $\ket{0}$ and perform a controlled-NOT gate operation so that $\ket{00}\rightarrow\ket{01}$ and $\ket{10}\rightarrow\ket{10}$. In this way, $\ket{\psi_{\Omega,2}}_{\text{ref}}$ can be prepared from the Dicke state
with an additional layer of two-qubit gates. The quantum circuit to prepare $\ket{\psi_{\Omega,2}}_{\text{ref}}$ from the Dicke state $\ket{D_1^{N/2}}$ is shown in Fig.~\ref{fig:D-to-ini}.

As another example, the input state of $\ket{qq}$ can be prepared in the following way. Since $\ket{qq}$ is a superposition of configurations of two fermions in the odd sites while the even sites are empty, the input state is then a Dicke state $\ket{D^{N/2}_2}$ on the odd sites with all even sites at the state $\ket{1}$. Input states for other hadrons can be constructed in a similar fashion.

\subsubsection{Constructing $U(\theta)$ with QAOA}

The operator $U(\theta)$ can be constructed by the quantum alternating operator ansatz~(QAOA). The Hamiltonian is split as $H=H_1+H_2+ \dots + H_M$, where $M\ge2$, with every $H_i$ inheriting the symmetries of $H$ and $[H_i,H_{i+1}]\not=0$. Then $U(\theta)$ is given by
\bea
U(\theta) \equiv \prod_{i=1}^p \prod_{j=1}^M
\exp(i\,\theta_{ij}H_j)\,.
\eea
Because every $H_i$ inherits the symmetries of $H$, the time evolution $\exp(i\theta_{ij} H_j)$ 
preserves the quantum numbers of the input reference state. The operator $\exp(i\theta_{ij} H_j)$ can be regarded as a rotation in the high-dimensional Hilbert space
about the axis $H_i$. The larger the $M$, the more 
choices of rotational axes we have. Furthermore, since $[H_i,H_{i+1}]$ 
does not vanish, successive rotations about $H_i$ and $H_{i+1}$ are equivalent to a rotation about a new axis. 
Therefore, the larger the values of $p$ and $M$, the better are the true hadronic states approximated by the trial states after the optimization. 
It should be noted that the required size of $p$ depends on the value of the bare coupling $g$.
If $g$ is small, say $g\sim 0.1$, $p$ can be chosen to be $N$; while if $g\sim 0.5$, $p$ can be chosen to be $N/2$.
 Optimization is achieved by minimizing the cost function.

In our case of the (1+1)D 1-flavor NJL model with staggered fermions, after the Jordan-Wigner transformation, 
the original Hamiltonian $H$ is split as $H=H_1+H_2+H_3+H_4$, with 
\bea
H_1 =& \sum_{n={\rm even}}^{\frac{N}{2}-1}
\frac{1}{4}\left(
\sigma_{n}^1 \sigma_{n+1}^2
- \sigma_{n}^2 \sigma_{n+1}^1
\right), \nn
H_2 =&   \sum_{n={\rm even}}^{\frac{N}{2}-1} \, \frac{g}{2}\, 
\sigma_{n}^3\sigma_{n+1}^3
 \,,  \nn
H_3 =& H_1(n={\rm even} \to n = {\rm odd}) \, \nn
& + \frac{1}{4}  \Xi^3_{N-1}\left(
\sigma_{N-1}^2 \sigma_{0}^1
- \sigma_{N-1}^1 \sigma_{0}^2 
\right)  \,, \nn
H_4 =& \sum_{n=0}^{\frac{N}{2}-1} \frac{m_q}{2}(-1)^n(I - \sigma_{n}^3)
- \frac{g}{2}(I - \sigma_{n}^3)\,.
\eea

We will consider the lowest-lying scalar meson state, i.e. the lightest hadron state $|h\rangle$ with the same quantum numbers as the vacuum.
\footnote{Note that in reality the lightest hadron in QCD has different parity from the vacuum.} The input reference states are as in eq.~(\ref{eq:NJLref}).

\subsection{Evaluation of the correlator}
For evaluation of the correlator, as depicted in Fig.~\ref{corr-circuit}, with the help of an ancillary qubit we measure the correlation function
\bea\label{corr}
	S_{mn}(t)=\bra{\Omega}e^{iHt}
	\Xi^3_m \sigma^i_m e^{-iHt}\Xi^3_n\sigma^j_n \ket{h}\,,
\eea
of which $\tilde\phi_h(z)$ in eq.~(\ref{eq:def_D}) can be written as a sum (see Ref.~\cite{Li:2021kcs} for details). 
In Fig.~\ref{corr-circuit}, the input state of the quantum circuit is taken as $\ket{\psi_{\Omega,1}}_{\rm ref}=\ket{0101,...,01}$. 
The quantum gate $D$ is implemented for preparation of the Dicke state for the reference hadron state: $D\ket{\psi_{\Omega,1}}_{\rm ref}=\ket{\psi_{h}}_{\rm ref}$. The unitary gate $U(\theta^*_\Omega)$ and $U(\theta^*_h)$ can help us prepare the vacuum and the hadronic state: $U(\theta^*_\Omega)\ket{\psi_{\Omega,1}}_{\rm ref}=\ket{\Omega}$, $U(\theta^*_h)\ket{\psi_{h}}_{\rm ref}=\ket{h}$. Specifically, if we want to prepare the $k$-th excited hadronic state with the same quantum numbers $l$ as the vacuum state, we have $\ket{\psi_{h}}_{\rm ref}=\ket{\psi_{\Omega,k}}_{\rm ref}$ and $\theta^{*}_{h}=\theta^{*}_{\Omega}$. Acting the controlled gates $D$, $U(\theta^*_\Omega)$, and $U(\theta^*_h)$ on the circuit will facilitate evaluation of the dynamical correlation function $\bra{\Omega}\mathcal{O}\ket{h}$, in which the bra state is different from the ket state. In short, when we act all the gates before the controlled $\Xi^3_n\sigma^j_n$ gate in the circuit, the state looks like $\frac{1}{\sqrt{2}}(\ket{0}\ket{\Omega}+\ket{1}\ket{h})$. After acting the controlled Pauli operators and the time evolution on the quantum circuit, we trace out the system and obtain the density matrix $\rho_A$ of the ancillary qubit. The $(\rho_{A})_{12}$ will have the form $\bra{\Omega}\mathcal{O}\ket{h}$ since the states $\ket{0}$ and $\ket{1}$ of the ancillary qubit are entangled with $\ket{\Omega}$ and $\ket{h}$ respectively.
 \begin{figure}[htbp]
    	\centering
    	\includegraphics[width=0.5 \textwidth]{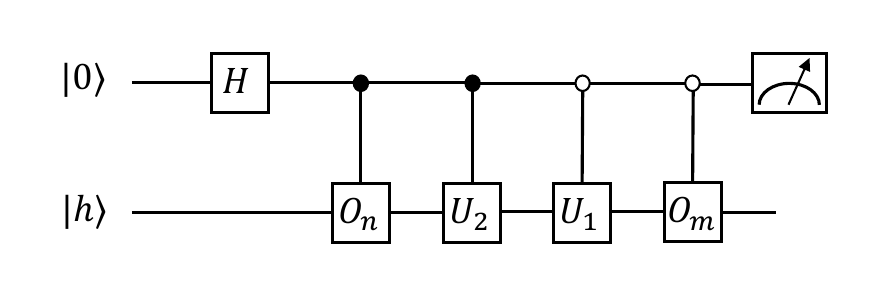}
    	\caption{Quantum circuit for the Hadamard test. The input state for the ancillary  qubit is $\ket{0}$ and the input state for the system is the hadronic state $\ket{h}$.}
    	\label{fig-HT}
    \end{figure}

It should be noted that, in the case of QCD, inclusion of the Wilson line will complicate the quantum circuit for the evaluation of the light-cone correlator. With the Wilson line, the light-cone correlator can still be expressed as a sum of $S_{mn}(t)=\bra{h}U_1(t)O_m U_2(t)O_n\ket{h}$, where 
$U_1$ and $U_2$ are unitary operators, and $O_m$ and $O_n$ are Hermitian operators. However, unlike the case without the Wilson line, now we have $U_1(t)U_2(t)\neq 1$. When $O_m$ and $O_n$ are the Pauli operators, which are unitary, $S_{mn}(t)$ can be viewed as an overlap between the states $U_1(t)O_m U_2(t)O_n\ket{h}$ and $\ket{h}$, which can be evaluated with the standard Hadamard test, the quantum circuit of which is shown in Fig.~\ref{fig-HT}.

\section{Results}
\label{res}
{The classical simulation of the quantum algorithm is performed on a desktop workstation with $16$ cores, using opensource packages QuSpin~\cite{quspin} and projectQ~\cite{Steiger2018projectqopensource}.
We perform the simulation of the LCDA for the lowest-lying scalar meson in the (1+1)D 1-flavor NJL model with $N=14$ qubits and different values of the bare coupling constant $g$ and hadron mass $m_h$. In practice, we first choose the values of the two free dimensionless parameters $m_qa$ and $g$, then from which we can determine the hadron state $\ket{h}$, its mass 
$m_h$ in units of $a^{-1}$, as well as its LCDA.  The values of $m_qa$ and $g$ are chosen in such a way that the condition $\frac{2\pi}{Na}<m_h<\frac{\pi}{a}$ is satisfied, so that the lattice size is bigger than the hadron size and the lattice spacing is smaller than the hadron size. We choose the phase of the hadronic state $|h\rangle$ such that $\phi_h(x)$ is a real function.
\begin{figure}[htbp]
	\centering
	\includegraphics[width=0.4\textwidth]{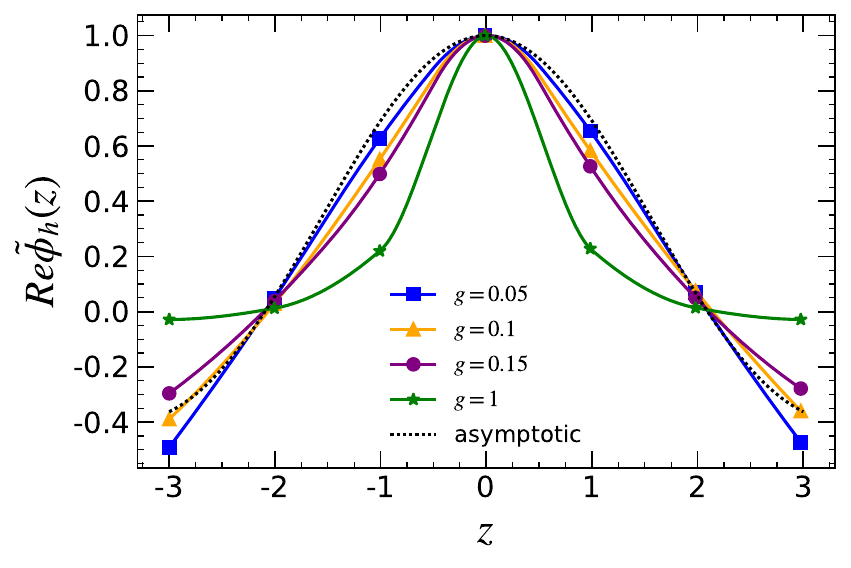}
\caption{Real part of $\tilde\phi_h(z)$ in the (1+1)D 1-flavor NJL model with $N=14$, $m_h=1.5a^{-1}$.}
	\label{fg:ReD}
\end{figure}
\begin{figure}[htbp]
	\centering
	\includegraphics[width=0.4\textwidth]{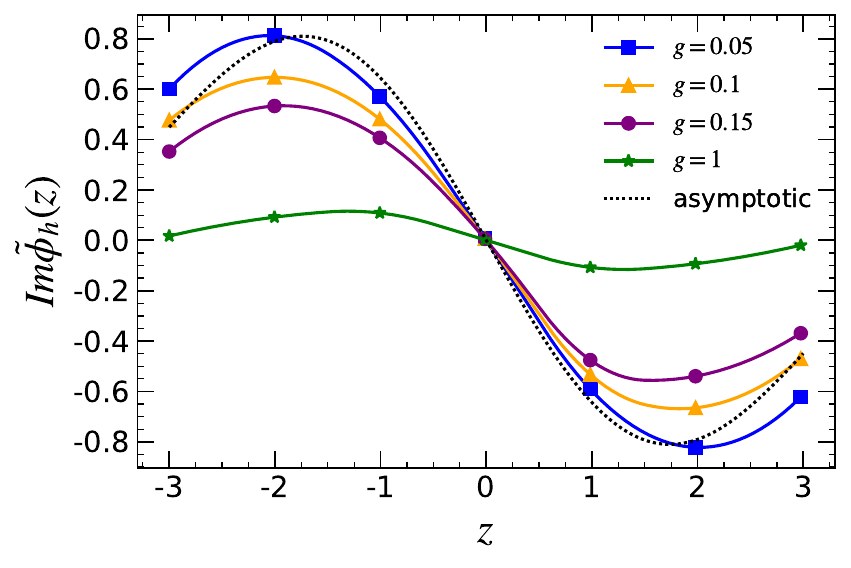}
\caption{Imaginary part of $\tilde\phi_h(z)$ in the (1+1)D 1-flavor NJL model with $N=14$, $m_h=1.5a^{-1}$.}
	\label{fg:ImD}
\end{figure}

We show in Figs.~\ref{fg:ReD} and~\ref{fg:ImD} the results for the real part and imaginary part of the LCDA in position space $\tilde\phi_h(z)$, respectively, with fixed value of $m_h=1.5a^{-1}$ and different values of $g$. We also show as dotted lines the inverse Fourier transform of the asymptotic form of $\phi_h(x)$ in QCD, namely $\phi_{\rm asymp}(x)=6x(1-x)$~\cite{Lepage:1980fj}. One notices that, as expected, the result from the quantum algorithm converges to the asymptotic LCDA as $g\to 0$, which mimics the  asymptotic behavior of LCDAs in QCD.    

Now we can evaluate the LCDA $\phi_h(x)$ by taking the Fourier transform of $\tilde\phi_h(z)$ as described in eq.~(\ref{eq:Fourier_decomp}). The final results are shown in Fig.~\ref{fg:gLCDA}, where the curves represent the results from exact diagonalization (ED), which are obtained from numerically diagonalizing the discretized NJL Hamiltonian, and the discrete open markers denote the results from quantum computing (QC) using classical hardware simulation. The excellent agreement between the results from the quantum algorithm and those from exact diagonalization justifies the designed quantum algorithm. Similar to Figs.~\ref{fg:ReD} and~\ref{fg:ImD}, we also show the asymptotic form of $\phi_h(x)$ in QCD for comparison. As expected, the peak of the LCDA gets narrower and converges to the asymptotic LCDA as $g\to 0$. Notice that there is non-vanishing but suppressed contributions in the nonphysical region ($x > 1$ or $x<0$). Such unphysical oscillations are caused by finite volume effects in the naively truncated Fourier transform, and are also commonly seen in lattice QCD calculations~\cite{Zhang:2020gaj}. 
We also check the dependence of the LCDA on the hadron mass $m_h$ in Fig.~\ref{fg:mhLCDA}. For this purpose, we take $m_h=1.3a^{-1},1.5a^{-1},1.7a^{-1}$, and fix the bare coupling constant $g=0.1$, with $N=14$. One can see that the peak of $\phi_h(x)$ gets narrower when the hadron mass increases. This is the expected behavior when the valence quark and antiquark become nonrelativistic, in which case the quark masses dominate the momenta of the quark and antiquark, while the relative momentum between the two becomes small. The behavior also agrees with the results from lattice QCD~\cite{Zhang:2020gaj}. 
We leave the extrapolation to the continuum limit as a follow-up work in the future.

\begin{figure}[htbp]
	\centering
	\includegraphics[width=0.45\textwidth]{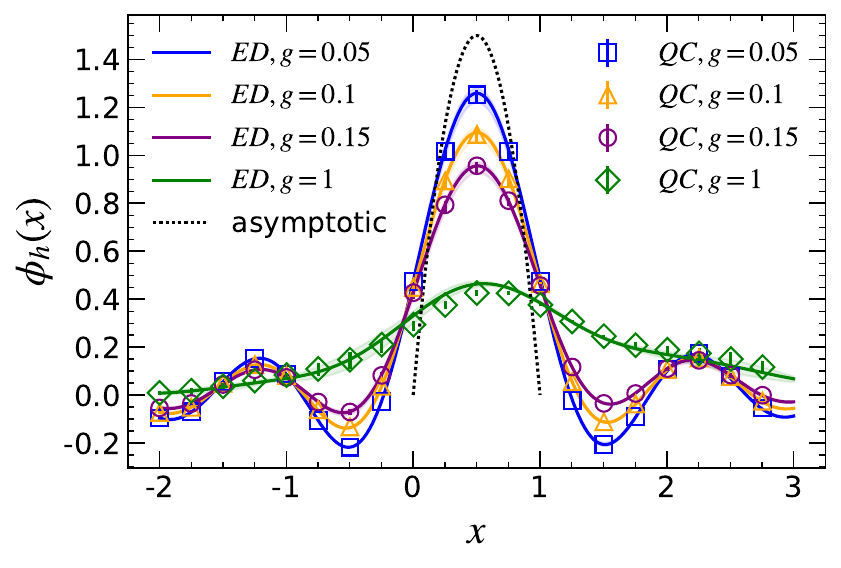}
\caption{LCDA in the (1+1)D 1-flavor NJL model with $N=14$, $m_h=1.5a^{-1}$.}
	\label{fg:gLCDA}
\end{figure}

\begin{figure}[htbp]
	\centering
	\includegraphics[width=0.45\textwidth]{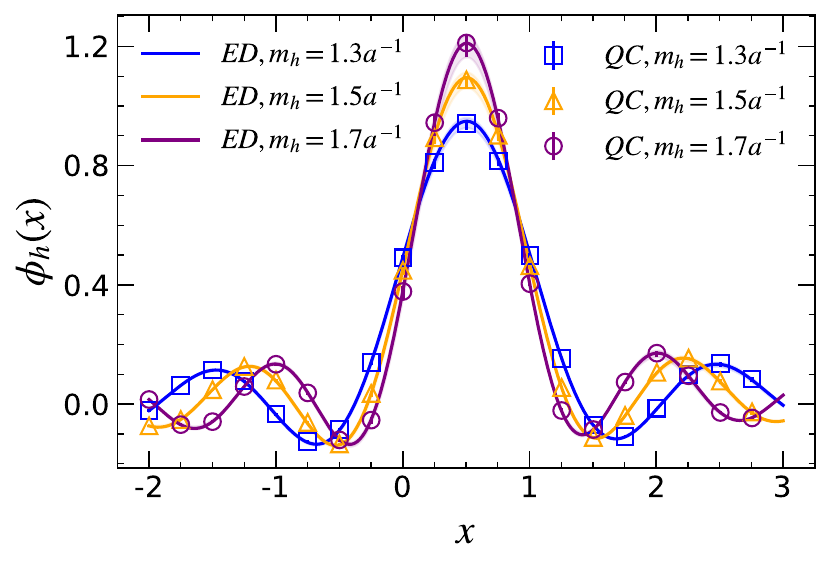}
\caption{Dependence of the LCDA on the hadron mass $m_h$ with fixed bare coupling $g=0.1$ in the (1+1)D 1-flavor NJL model.}
	\label{fg:mhLCDA}
\end{figure}

\section{Summary}
\label{sum}
In this study, we presented the first direct simulation of the light-cone distribution amplitude (LCDA) with a quantum algorithm on classical hardware. 
Using a quantum algorithm we proposed recently for the evaluation of the parton distribution functions, here we 
performed the simulation for the LCDA in the (1+1)-dimensional Nambu-Jona-Lasinio (NJL) model on classical hardware.
With $14$ qubits, our results from the quantum algorithm agree with those from exact diagonalization of the discretized NJL model.
Our results of the LCDA showed the expected dependence on the coupling constant and the hadron mass. 

The result presented in this study manifests the feasibility of using quantum computing to solve intrinsic difficulties in realizing  real-time dynamics with classical computing facilities. Meanwhile, it demonstrates that the recently proposed quantum computing framework for preparing  hadronic states and measuring  dynamical correlation functions is generally applicable.  The extension of the algorithm to other applications in high energy particle and nuclear physics can be expected.

~~~~~
\begin{acknowledgments}
This work is supported by the National Natural Science Foundation of China under Grant No. 12022512, No.~12035007 (H.X.), Grant No.~12175016 (X.L.), Grant No.12005065 (D.B.), Grant No. 12074180 (S.L.), and by the Guangdong Major Project of Basic and Applied Basic Research No. 2020B0301030008, the Key-Area Research and Development Program of GuangDong Province (Grant No. 2019B030330001), the Guangdong Basic and Applied Basic Research Fund No.2021A1515010317 (D.B.), the Key Project of Science and Technology of Guangzhou (Grant No. 2019050001), the National Special
Support Program for High-level Talents (X.L.).
\end{acknowledgments}




\begin{thebibliography}{10}

\bibitem{H1:2015ubc}
H.~Abramowicz \textit{et al.},
\newblock Eur. Phys. J. C \textbf{75}, no.12, 580 (2015)

\bibitem{AbdulKhalek:2021gbh}
R.~Abdul Khalek, A.~Accardi, J.~Adam, D.~Adamiak, W.~Akers, M.~Albaladejo, A.~Al-bataineh, M.~G.~Alexeev, F.~Ameli and P.~Antonioli, \textit{et al.}
Nucl. Phys. A \textbf{1026}, 122447 (2022)

\bibitem{EicCWP}
Cao, X. \textit{et al.}, { Nucl.
  Tech.\/}, {\bf 43}, 020001 (2020).

\bibitem{Anderle:2021wcy}
D.~P.~Anderle, V.~Bertone, X.~Cao, L.~Chang, N.~Chang, G.~Chen, X.~Chen, Z.~Chen, Z.~Cui and L.~Dai \textit{et al.},
Front. Phys. \textbf{16}, 64701 (2021).


\bibitem{Lepage:1980fj}
G.~P. Lepage and S.~J. Brodsky,
\newblock Phys. Rev. D {\bf 22}, 2157 (1980).

\bibitem{Efremov:1979qk}
A.~V. Efremov and A.~V. Radyushkin,
\newblock Phys. Lett. B {\bf 94}, 245 (1980).

\bibitem{Chernyak:1983ej}
V.~L. Chernyak and A.~R. Zhitnitsky,
\newblock Phys. Rep. {\bf 112}, 173 (1984).

\bibitem{Chernyak:1981zz}
V.~L. Chernyak and A.~R. Zhitnitsky,
\newblock Nucl. Phys. B {\bf 201}, 492 (1982),
\newblock [Erratum: Nucl.Phys.B 214, 547 (1983)].

\bibitem{Chernyak:1984bm}
V.~L. Chernyak and I.~R. Zhitnitsky,
\newblock Nucl. Phys. B {\bf 246}, 52 (1984).

\bibitem{King:1986wi}
I.~D. King and C.~T. Sachrajda,
\newblock Nucl. Phys. B {\bf 279}, 785 (1987).

\bibitem{Chernyak:1987nu}
V.~L. Chernyak, A.~A. Ogloblin, and I.~R. Zhitnitsky,
\newblock Yad. Fiz. {\bf 48}, 1410 (1988).

\bibitem{Chernyak:1987nv}
V.~L. Chernyak, A.~A. Ogloblin, and I.~R. Zhitnitsky,
\newblock Yad. Fiz. {\bf 48}, 1398 (1988).

\bibitem{Radyushkin:1990te}
A.~V. Radyushkin,
\newblock Nucl. Phys. A {\bf 532}, 141 (1991).

\bibitem{RuizArriola:2006jge}
E.~Ruiz~Arriola and W.~Broniowski,
\newblock Phys. Rev. D {\bf 74}, 034008 (2006), arXiv:hep-ph/0605318.

\bibitem{RuizArriola:2002bp}
E.~Ruiz~Arriola and W.~Broniowski,
\newblock Phys. Rev. D {\bf 66}, 094016 (2002), arXiv:hep-ph/0207266.

\bibitem{Chang:2013pq}
L.~Chang \textit{ et~al.},
\newblock Phys. Rev. Lett. {\bf 110}, 132001 (2013), arXiv:1301.0324.

\bibitem{Agaev:2012tm}
S.~S. Agaev, V.~M. Braun, N.~Offen, and F.~A. Porkert,
\newblock Phys. Rev. D {\bf 86}, 077504 (2012), arXiv:1206.3968.



\bibitem{Brodsky:2006uqa}
S.~J.~Brodsky and G.~F.~de Teramond,
\newblock Phys. Rev. Lett. \textbf{96}, 201601 (2006)

\bibitem{Vary:2009gt}
J.~P.~Vary, H.~Honkanen, J.~Li, P.~Maris, S.~J.~Brodsky, A.~Harindranath, G.~F.~de Teramond, P.~Sternberg, E.~G.~Ng and C.~Yang,
\newblock Phys. Rev. C \textbf{81}, 035205 (2010)

\bibitem{Brodsky:2014yha}
S.~J.~Brodsky, G.~F.~de Teramond, H.~G.~Dosch and J.~Erlich,
\newblock Phys. Rep. \textbf{584}, 1-105 (2015)

\bibitem{Xu:2021wwj}
S.~Xu \textit{et al.} [BLFQ],
Phys. Rev. D \textbf{104}, no.9, 094036 (2021)




\bibitem{Ma:2006hc}
J.~P. Ma and Z.~G. Si,
\newblock Phys. Lett. B {\bf 647}, 419 (2007), arXiv:hep-ph/0608221.

\bibitem{Jia:2008ep}
Y.~Jia and D.~Yang,
\newblock Nucl. Phys. B {\bf 814}, 217 (2009), arXiv:0812.1965.

\bibitem{Gockeler:2008xv}
M.~Gockeler \textit{ et~al.},
\newblock Phys. Rev. Lett. {\bf 101}, 112002 (2008), arXiv:0804.1877.

\bibitem{QCDSF:2008qtn}
V.~M. Braun \textit{ et~al.},
\newblock Phys. Rev. D {\bf 79}, 034504 (2009), arXiv:0811.2712.

\bibitem{Braun:2014wpa}
V.~M. Braun \textit{ et~al.},
\newblock Phys. Rev. D {\bf 89}, 094511 (2014), arXiv:1403.4189.

\bibitem{Bali:2015ykx}
G.~S. Bali \textit{ et~al.},
\newblock J. High Energ. Phys. {\bf 02}, 070 (2016), arXiv:1512.02050.

\bibitem{RQCD:2019osh}
G.~S. Bali \textit{ et~al.},
\newblock J. High Energ. Phys. {\bf 08}, 065 (2019), arXiv:1903.08038.

\bibitem{Zhang:2020gaj}
R.~Zhang, C.~Honkala, H.-W. Lin, and J.-W. Chen,
\newblock Phys. Rev. D {\bf 102}, 094519 (2020), arXiv:2005.13955.

\bibitem{Zhang:2017bzy}
J.-H. Zhang, J.-W. Chen, X.~Ji, L.~Jin, and H.-W. Lin,
\newblock Phys. Rev. D {\bf 95}, 094514 (2017), arXiv:1702.00008.

\bibitem{Zhang:2017zfe}
J.-H. Zhang \textit{ et~al.},
\newblock Nucl. Phys. B {\bf 939}, 429 (2019), arXiv:1712.10025.

\bibitem{Hua:2020gnw}
J.~Hua \textit{ et~al.},
\newblock Phys. Rev. Lett. {\bf 127}, 062002 (2021), arXiv:2011.09788.

\bibitem{Hua:2022kcm}
J.~Hua \textit{ et~al.}, arXiv:2201.09173.

\bibitem{Alexandru:2016gsd}
A.~Alexandru, G.~Basar, P.~F. Bedaque, S.~Vartak, and N.~C. Warrington,
\newblock Phys. Rev. Lett. {\bf 117}, 081602 (2016), arXiv:1605.08040.

\bibitem{Arute:2019zxq}
F.~Arute \textit{ et~al.},
\newblock Nature {\bf 574}, 505 (2019), arXiv:1910.11333.

\bibitem{Cloet:2019wre}
M.~R. Dietrich \textit{ et~al.},
\newblock {Opportunities for Nuclear Physics and Quantum Information Science}, arXiv:1903.05453.

\bibitem{Zhang:2020uqo}
D.-B. Zhang, H.~Xing, H.~Yan, E.~Wang, and S.-L. Zhu,
\newblock Chin. Phys. B {\bf 30}, 020306 (2021), arXiv:2011.01431.

\bibitem{Bauer:2022hpo}
C.~W. Bauer \textit{ et~al.}, arXiv:2204.03381.

\bibitem{Jordan:2012xnu}
S.~P. Jordan, K.~S.~M. Lee, and J.~Preskill,
\newblock Science {\bf 336}, 1130 (2012), arXiv:1111.3633.

\bibitem{Jordan:2011ci}
S.~P. Jordan, K.~S.~M. Lee, and J.~Preskill,
\newblock Quant. Inf. Comput. {\bf 14}, 1014 (2014), arXiv:1112.4833.

\bibitem{Jordan:2014tma}
S.~P. Jordan, K.~S.~M. Lee, and J.~Preskill, arXiv:1404.7115.

\bibitem{Klco:2018zqz}
N.~Klco and M.~J. Savage,
\newblock Phys. Rev. A {\bf 99}, 052335 (2019), arXiv:1808.10378.

\bibitem{Dumitrescu:2018njn}
E.~F. Dumitrescu \textit{ et~al.},
\newblock Phys. Rev. Lett. {\bf 120}, 210501 (2018), arXiv:1801.03897.

\bibitem{Lu:2018pjk}
H.-H. Lu \textit{ et~al.},
\newblock Phys. Rev. A {\bf 100}, 012320 (2019), arXiv:1810.03959.

\bibitem{Lamm:2019uyc}
H.~Lamm, S.~Lawrence, and Y.~Yamauchi,
\newblock Phys. Rev. Res. {\bf 2}, 013272 (2020), arXiv:1908.10439.

\bibitem{Mueller:2019qqj}
N.~Mueller, A.~Tarasov, and R.~Venugopalan,
\newblock Phys. Rev. D {\bf 102}, 016007 (2020), arXiv:1908.07051.

\bibitem{Roggero:2019myu}
A.~Roggero, A.~C.~Y. Li, J.~Carlson, R.~Gupta, and G.~N. Perdue,
\newblock Phys. Rev. D {\bf 101}, 074038 (2020), arXiv:1911.06368.

\bibitem{Echevarria:2020wct}
M.~G. Echevarria, I.~L. Egusquiza, E.~Rico, and G.~Schnell, arXiv:2011.01275.

\bibitem{Bauer:2021gup}
C.~W. Bauer, M.~Freytsis, and B.~Nachman, arXiv:2102.05044.

\bibitem{Atas:2021ext}
Y.~Y. Atas \textit{ et~al.},
\newblock Nature Commun. {\bf 12}, 6499 (2021), arXiv:2102.08920.

\bibitem{Li:2021kcs}
T.~Li \textit{ et~al.},
\newblock Phys. Rev. D {\bf 105}, L111502 (2022), arXiv:2106.03865.

\bibitem{Kreshchuk:2020kcz}
M.~Kreshchuk, S.~Jia, W.~M.~Kirby, G.~Goldstein, J.~P.~Vary and P.~J.~Love,
Entropy \textbf{23}, no.5, 597 (2021)




\bibitem{Gallimore:2022hai}
D.~Gallimore and J.~Liao, arXiv:2202.03333.

\bibitem{Martinez:2016yna}
E.~A. Martinez \textit{ et~al.},
\newblock Nature {\bf 534}, 516 (2016), arXiv:1605.04570.

\bibitem{Hu:2019hrf}
Z.~Hu, R.~Xia, and S.~Kais,
\newblock Sci. Rep. {\bf 10}, 3301 (2020), arXiv:1904.00910.

\bibitem{Bauer:2019qxa}
C.~W. Bauer, W.~A. de~Jong, B.~Nachman, and D.~Provasoli,
\newblock Phys. Rev. Lett. {\bf 126}, 062001 (2021), arXiv:1904.03196.

\bibitem{DeJong:2020riy}
W.~A. De~Jong \textit{ et~al.},
\newblock Phys. Rev. D {\bf 104}, 051501 (2021), arXiv:2010.03571.

\bibitem{Zhou:2021kdl}
Z.-Y. Zhou \textit{ et~al.}, arXiv:2107.13563.

\bibitem{deJong:2021wsd}
W.~A. de~Jong \textit{ et~al.}, arXiv:2106.08394.

\bibitem{Williams:2021lvr}
S.~Williams, S.~Malik, M.~Spannowsky, and K.~Bepari, arXiv:2109.13975.

\bibitem{Atas:2022dqm}
Y.~Y. Atas \textit{ et~al.}, arXiv:2207.03473.

\bibitem{Yao:2022eqm}
X.~Yao, arXiv:2205.07902.

\bibitem{Czajka:2021yll}
A.~M. Czajka, Z.-B. Kang, H.~Ma, and F.~Zhao, arXiv:2112.03944.

\bibitem{Xie:2022jgj}
X.-D. Xie \textit{ et~al.}, arXiv:2205.12767.

\bibitem{Banuls:2019bmf}
M.~C. Ba\~nuls \textit{ et~al.},
\newblock Eur. Phys. J. D {\bf 74}, 165 (2020), arXiv:1911.00003.

\bibitem{Nambu:1961tp}
Y.~Nambu and G.~Jona-Lasinio,
\newblock Phys. Rev. {\bf 122}, 345 (1961).

\bibitem{Nambu:1961fr}
Y.~Nambu and G.~Jona-Lasinio,
\newblock Phys. Rev. {\bf 124}, 246 (1961).

\bibitem{Gross:1974jv}
D.~J. Gross and A.~Neveu,
\newblock Phys. Rev. D {\bf 10}, 3235 (1974).

\bibitem{backens_shnirman_makhlin_2019}
S.~Backens, A.~Shnirman, and Y.~Makhlin,
\newblock Sci. Rep. {\bf 9} (2019).

\bibitem{10.1007/978-3-030-25027-0_9}
A.~B{\"a}rtschi and S.~Eidenbenz,
\newblock Deterministic preparation of dicke states, in L.A. Gasienice, J. Jansson, and C. Levcopoulos, eds.,
\newblock \textit{ Fundamentals of Computation Theory} (Springer, Cham,2019), pp. 126--139.

\bibitem{nielsen_chuang_2010}
M. A. Nielsen, and I.L. Chuang,
\newblock \textit{ Quantum Computation and Quantum
Information} {(Cambridge University Press, Cambridge, 2010).}

\bibitem{quspin}
P.~Weinberg and M.~Bukov,
\newblock SciPost Phys. {\bf 2}, 003 (2017).

\bibitem{Steiger2018projectqopensource}
D.~S. Steiger, T.~H{\"{a}}ner, and M.~Troyer,
\newblock {Quantum} {\bf 2}, 49 (2018).



\end{thebibliography}
\end{document}